\documentstyle[sprocl]{article}
\begin{document}
\title{Solution of $a_2/a_1$ sign problem in heavy meson decays:\\
 nonfactorizable soft gluons in nonleptonic heavy meson decays}
\author{B.Tseng and Hsiang-nan Li}
\address{
Department of Physics, National Cheng-Kung University,
Tainan, Taiwan}
\maketitle
\abstracts{
For the first time, a solution to the $a_2/a_1$ problem in heavy meson 
decays is proposed, based on PQCD factorization theorems. The 
point is to include QCD evolution effects among 
various characteristic scales involved in decay processes.
}
\section{Introduction}
 
Nonleptonic weak decays are notoriously difficult to handle.
Approximations are needed to simplify the analyses.
The most widely adopted approach to exclusive nonleptonic
heavy meson decays is the Bauer-Stech-Wirbel (BSW) model \cite{BSW} based
on the factorization hypothesis, in which decay rates are expressed in
terms of various hadronic transition form factors multiplied by some
coefficients. The coefficients corresponding to
external $W$ boson emission and to internal $W$ boson emission
are $a_1=c_1+c_2/N_c$ and $a_2=c_2+c_1/N_c$, respectively, 
$c_i$ being the Wilson coefficients. 
The nonfactorizable contributions which cannot be expressed in terms of 
hadronic form factors, and the nonspectator contributions from $W$
boson exchanges are neglected. In this way the BSW method avoids
complicated QCD dynamics. In this approach  the fierz-tranformation term
is treated as a free parameter $\xi$ (=$1/N_c$) and it is found that over 20 
decay 
modes can be sucessfully explained if $\xi$=0 which means that $N_c$ goes to 
infinity and the large $N_c$ approach is confirmed\cite{BG}. 
To understand the sucesses of this large $N_c$ approach, 
nonfactorizable 
contributions $\chi$ are included, and the effective coefficients $a_1$ and 
$a_2$ become
\begin{equation}
a^{\rm eff}_1=c_1+c_2\left(\frac{1}{N_c}+\chi_1\right)\;,\;\;\;\;
a^{\rm eff}_2=c_2+c_1\left(\frac{1}{N_c}+\chi_2\right)\;.
\end{equation}
People have calculated these nonfactorizable
contributions by using QCD sum rules\cite{BS87} and found that the 
parameters $\chi$ are negative for charm decays, canceling the 
color-suppressed term $1/N_c$. Hence, the famous " rule of discarding the 
$1/N_c$ corrections" \cite{BS87} found its
dynamical origin in QCD.

Performing a similar BSW analysis  for B meson decays\cite{CT95}, it is 
found that
$a_2/a_1$ is positive, in contrast to the negative $a_2/a_1$ for 
the charmed meson. The values $c_{1}(M_b)\approx 1.12$ and 
$c_{2}(M_b)\approx -0.26$ imply that a  naive extension of 
the large $N_c$ approach to bottom decays\cite{NS97} will lead to a negative 
$a_2/a_1$ , which is in conflict to experimental data. 
If nonfactorizable contributions are included into $a_2$,
we will need  positive $\chi$ to explain the data.
Unfortunally, all the existing calculations give  negative
nonfactorizable contributions\cite{BS93}. Up to now, it is still 
unclear what is wrong with these calculations and with the 
QCD-sum-rule method.
This $a_2/a_1$ sign problem 
provides a challenging subject for the QCD sum rule method.


\section{PQCD formalism for nonleptonic heavy meson deacys}

>From the viewpoint of PQCD theory, nonleptonic weak decays are a
three-scale problem. The involved scales are $M_W$, the W boson 
mass, t, the typical scale of the meson decay and $\Lambda_{QCD}$, 
which characterizes the hadronic size. Combining 
the effective field theory and the PQCD factorization theorem, a
three-scale factorization formula has been proposed and applied to the $B 
\rightarrow J/\Psi K^*$ decay with success \cite{CL}.
We assume that a parton carries small amount of transverse momenta 
$k_{T}$, which serve as an infrared regulator for radiative corrections.
The factorization formula for the decay amplitude is  
written as
\begin{equation}
{\cal M}=H_r(M_W,\mu)\otimes H(t,\mu)\otimes \phi(b,\mu)\otimes
U(b,\mu)\;,
\label{hpu}
\end{equation}
in the impact parameter b space conjugate to $k_{T}$, where $\otimes$ 
represents a 
convolution relation, since variables $t$ and $b$ will be integrated out at 
last.
According to the virtuality of gluon momenta, QCD dynamics are seperated 
into a "harder" function $H_r$, a hard decay amplitude $H$ and soft parts
$\phi$ and $U$. 
The reduciable soft gluons are absorbed into $\phi(b,\mu)$, 
while the irreduciable ones into $U(b,\mu)$.
 
Let $\gamma_{H_r}$, $\gamma_\phi$, and
$\gamma_U$ be the anomalous dimensions of $H_r$, $\phi$, and $U$,
respectively. The anomalous dimension of $H$ is then
$\gamma_H=-(\gamma_{H_r}+\gamma_\phi+\gamma_U)$, because a decay
amplitude is ultraviolet finite. A RG treatment of
Eq.~(\ref{hpu}) leads to
\begin{eqnarray}
{\cal M}&=&H_r(M_W,M_W)\otimes H(t,t)\otimes \phi(b,1/b)\otimes U(b,1/b)
\nonumber \\
& &\otimes
\exp\left[\int_{t}^{M_W}\frac{d{\bar\mu}}
{\bar\mu}\gamma_{H_r}(\alpha_s({\bar\mu}))\right]
\nonumber\\
& &\otimes
\exp\left[-\int_{1/b}^t\frac{d{\bar\mu}}
{\bar\mu}\left[\gamma_\phi(\alpha_s({\bar\mu}))
+\gamma_U(\alpha_s({\bar\mu}))\right]\right]\;,
\label{main}
\end{eqnarray}
which becomes explicitly $\mu$-independent. The two exponentials,
describing the two-stage evolutions from $M_W$ to $t$ and from $t$ to
$1/b$,
are the consequence of the summation of large $\ln(M_W/t)$ and
$\ln(tb)$. The first exponential can be easily identified as the Wilson
coefficient $c(t)$. 

Equation (\ref{main}) sums only the single logarithms. In fact, there
exist
the double logarithms $\ln^2(P^+b)$ in the meson wave function $\phi$,
which
arise from the overlap of collinear and soft divergences,
$P^+$ being the largest light-cone component of the meson momentum.  
Hence, $\phi(b,\mu)$ in Eq.~(\ref{hpu}) should be replaced by
\begin{equation}
\phi(P^+,b,\mu)=\phi(b,\mu)\exp[-s(P^+,b)]\;,
\label{sdc}
\end{equation}
where the exponential $e^{-s}$, the so-called Sudakov form factor, comes
from the resummation of the double logarithms \cite{LY1,L1}. 
We shall approximate $H_r(M_W,M_W)$ by
its lowest-order expression $H^{(0)}_r=1$, and 
evaluate $H(t,t)$
perturbatively, since all the large logarithms have been grouped into the
exponents. 
For simplicity, we set the nonperturbative initial condition
$U(b,1/b)$ of the RG evolution to unity due to the strong suppression 
from the Sudakov form factor at
large $b$, and neglect the $b$ dependence in another
nonperturbative initial condition $\phi(b,1/b)$.

We need to calculate  
$\gamma_U$, which is complicated due to the nonabelian nature
of color exchanges.
The color structure of soft gluon exchanges can be analyzed as follows. 
If a soft  gluon  crosses the hard gluon vertex, the color structure is 
given by  
\begin{eqnarray}
& &(T^a)_{b_1 b^{\prime}_1}(T^H)_{b^{\prime}_1 b_2}
(T^a)_{a_2 a^{\prime}_2}(T^H)_{a^{\prime}_2 a_1}
\nonumber\\
& &=\frac{1}{2}(T^H)_{a_2 b_2}(T^H)_{b_1 a_1}-  
{1 \over 2N_c}(T^H)_{b_1 b_2}(T^H)_{a_2 a_1}\;,
\end{eqnarray}
where $T^a$ $(T^H)$ is the color matrix
associated with the soft (hard) gluon. Contracted with
$\delta_{a_1 b_1} \delta_{a_2 b_2}$ from the color-singlet initial- and
final-state mesons, the first term, denoting an octet contribution,
diminishes. The second term leads to a factor $-1/(2 N_c)$ without
changing
the original tree-level color flow. If a soft gluon does not cross the
hard gluon vertex, it introduces a factor $T^aT^a={\cal C}_F=4/3$.

\section{$B^-\to D^0\pi^-$ and  $D^+\to \bar{K}^0\pi^+$: solution of 
$a_2/a_1$ sign problem}

Now we are ready to see how to solve this $a_2/a_1$ sign problem within 
our PQCD approach. Since the analyses of  $D^+\to \bar{K}^0\pi^+$ is 
similar to
that of $B^-\to D^0\pi^-$(except for revelent kinematics), we will focus on 
$B^-\to D^0\pi^-$.

The decay amplitude is given by
\begin{eqnarray}
{\cal M}=f_\pi[(1+r)\xi_+-(1-r)\xi_-]+
f_D\xi_i+{\cal M}_e+{\cal M}_i\;,
\label{M1}
\end{eqnarray}
$f_{D}$ and $f_\pi$ being the $D$ meson and pion decay constant,
respectively.
The form factor $\xi_{\pm}$ come from the factorizable external $W$ 
contributions and $\xi_{i}$ from the internal $W$ contributions. 
The amplitudes $M_e$ and $M_i$ denote the corresponding nonfactorizable 
contributions. After including the soft gluons, $\xi$ becomes 
nonfactorizable. 
The nonfactorizable contribution can be extracted by 
subtracting its (lowest order) factorizable part. This is one of the 
key points to the solution of the $a_2/a_1$ sign problem.
We refer readers to \cite{LT97} for more details.

After including the soft gluons, the factorizable external $W$-emission
contributon $\chi^{(f)}_e$ and the factorizable internal $W$-emission
contributon $\chi^{(f)}_i$ to the decay amplitude ${\cal M}$ in
Eq.~(\ref{M1}) should be identified as
$\chi^{(f)}_e=f_\pi[(1+r)\xi_+-(1-r)\xi_-]\;$,
$\chi^{(f)}_i=f_{D(K)}\xi_i|_{S_U=0}\;$,
where $\xi_{i}|_{S_U=0}$ denotes the lowest order internal $W$-emission
. The nonfactorizable external $W$-emission contributon   
$\chi^{(nf)}_e$ and the nonfactorizable internal $W$-emission contributon
$\chi^{(nf)}_i$  are then
$\chi^{(nf)}_e={\cal M}_e\;$,
$\chi^{(nf)}_i=f_{D(K)}(\xi_{i}-\xi_{i}|_{S_U=0})+{\cal M}_i\;$.
 
The results of the various form factors
and amplitudes for the decays $B^-\to D^{0}\pi^-$ and
$D^+\to {\bar K}^{0}\pi^+$ are exhibited in the following Table(unit = 
$10^{-3}$ GeV). 

\vskip 0.1cm
\[\begin{array}{ccccc}\hline\hline
B\to D\pi&{\rm external}\;W&f_D\xi_{i}&{\cal M}_e&{\cal M}_i\\
         &{\rm (factorizable)}&       &          &          \\
\hline
S_U=0    &106.5 &2.5 &-5.8+19.8i &18.5-12.5i\\
S_U\not=0 &108.5 &2.6 &-5.8+20.0i &18.8-11.0i\\
\hline \hline
D\to {\bar K}\pi&{\rm external}\;W&f_K\xi_{i}&{\cal M}_e&{\cal M}_i\\
         &{\rm (factorizable)}&       &          &          \\
\hline
S_U=0    &267.0 &-21.3 &-18.8+19.0i &37.8-36.7i\\
S_U\not=0 &1075.0 &-529.3 &-18.5+13.6i &36.5-30.3i\\
\hline\hline
         &\chi^{(f)}_e &\chi^{(f)}_i &\chi^{(nf)}_e &\chi^{(nf)}_i \\
\hline
B\to D\pi & 108.5 & 2.5 & -5.8+20.0i & 18.9-11.0i \\
D\to {\bar K}\pi & 1075.0 & -21.3 & -18.5+13.6i & -471.5-30.3i \\
\hline\hline

\end{array}\]

The rows entitled by
$S_U\not =0$, whose values match the data, are derived with the soft
corrections taken into account. Those by $S_U=0$ only help to extract 
the  
nonfactorizable contributions $\chi^{(nf)}$, and to investigate the
importance of the soft corrections. 
The exponentials $e^{-S_U}$, 
which basically act as
enhancing factors, then amplify the contributions from the region with
smaller $t$, where the Wilson coefficients are larger. Therefore, the soft
gluon effects are more important in charm decays as show in the above Table .

The factorizable external $W$-emission contributions are positive in both
the $B$ and $D$ meson decays, and their magnitudes increase, after including
the soft corrections with $\gamma_U<0$. $\xi_{i}$ changes sign, since
the Wilson coefficient $c_2$  becomes so negative, when evolving
from the characteristic scale of the $B$ meson decay to that of the $D$ 
meson decay, that it overcomes the positive $c_1/N_c$. Their
magnitudes also increase because of $\gamma_U<0$.
The real parts of the nonfactorizable amplitudes ${\cal M}_e$ (${\cal M}_i$)
are always negative (positive) due to the negative $c_2$ (positive   
$c_1/N_c$) in the $B$ and $D$ meson decays. 
>From the table,
we have $Re(\chi^{(nf)}_i)=+0.0189$ GeV for the $B$ meson decay and
$Re(\chi^{(nf)}_i)=-0.4715$ GeV for the $D$ meson decay.
It is easy to attribute the sign change of $\chi^{(nf)}_i$
to the stronger enhancement of $\xi_i$ by the soft corrections in
charm decays.

\section{Summary}

In this talk, I describe a solution to the $a_2/a_1$ sign problem.
The nonfactorizable contributions in
nonleptonic heavy meson decays are carefully identified in our formalism, 
and found to be positive for bottom decays and negative for charm decays
with the inclusion of soft gluons.

\noindent {\bf ACKNOWLEDGMENTS}

B. Tseng would like to thank Profs. K.Hagiwara and A. Ali for valuable 
discussions during the workshop.
This work was supported in part by the 
National Science Council of ROC under Contract Nos. NSC87-2112-M-006-018.

\section*{References}
\newcommand{\bi}{\bibitem}

\end{document}